# Dynamics of fragmentation and multiple vacancy generation in irradiated single-walled carbon nanotubes

Sumera Javeed [a], Sumaira Zeeshan [a], Shoaib Ahmad [b,c,*]

[a] PINSTECH & PIEAS P.O. Nilore, Islamabad, Pakistan
[b] Government College University (GCU), CASP, Church Road, 54000 Lahore, Pakistan
[c] National Centre for Physics, Quaid-i-Azam University Campus, Shahdara Valley, Islamabad, 44000, Pakistan

[*] Email: sahmad.ncp@gmail.com

**Abstract**
The results from mass spectrometry of clusters sputtered from $Cs^+$ irradiated single-walled carbon nano-tubes (SWCNTs) as a function of energy and dose identify the nature of the resulting damage in the form of multiple vacancy generation. For pristine SWCNTs at all $Cs^+$ energies, $C_2$ is the most dominant species, followed by $C_3$, $C_4$ and $C_1$. The experiments were performed in three stages: in the first stage, $Cs^+$ energy $E(Cs^+)$ was varied. During the second stage, the nanotubes were irradiated continuously at $E(Cs^+) = 5$ keV for 1,800 s. Afterwards, the entire sequence of irradiation energies was repeated to differentiate between the fragmentation patterns of the pristine and of heavily irradiated SWCNTs. The sputtering and normalized yields identify the quantitative and relative extent of the ion-induced damage by creating double, triple and quadruple vacancies; the single vacancies are least favored. Sputtering from the heavily irradiated SWCNTs occurs not only from the damaged and fragmented nanotubes, but also from the inter-nanotube structures that are grown due to the accumulation of the sputtered clusters. Similar irradiation experiments were performed with the multi-walled carbon nanotubes; the results confirmed the dominant $C_2$ followed by $C_3$, $C_4$ and $C_1$.

## 1. Introduction

The discovery and rapid evolution of carbon nanotubes (CNTs) has opened new horizons and stimulated research and development in the field of nanotechnology. The synthesis of carbon nano-tubes was first reported by Iijima [1] and in gram quantities by Ebbesen [2] using a variant of the standard arc-discharge technique for fullerene synthesis under helium atmosphere. CNTs have a wide variety of interesting properties, as well as many potential applications and can be used in nanometer scale devices and new composite materials [3,4]. In CNTs, the collision of an energetic particle with carbon atoms results in formation of a vacancy (single- or multiple) and a number of primary knock-on atoms which, if their energy is high enough, can leave the nanotube or displace other atoms within. Much work has been conducted on the irradiation of carbon nano structures using combinations of simulations and experiments. Irradiation with ions introduces a number of structural defects, with the most common being vacancies in single- and multi-walled carbon nanotubes. Molecular dynamics has been employed to simulate the irradiation of CNTs with various noble-gas ions, calculate the ion ranges as a function of ion energy [5–7] and study the ion impacts on nanotubes lying on different substrates. Defect production depends on the type of the substrate, and the damage induced is higher for the metallic heavy-atom substrates than for the light-atom ones. The coaxial nanotube network of multi-walled carbon nanotubes (MWCNTs) appears to anneal the localized damage due to defect migration and the saturation of the resulting dangling-bonds [8]. The ratio of single to double vacancies is minimal at ion energies of approximately 0.5 keV and becomes saturated towards a constant value at high ion energies [9]. Ion irradiation results in the welding of crossed nanotubes that are suspended or deposited on substrates [10]. The chemical functionalization of the nanotubes and formation of defects on the nanotube walls was predicted using a combination of computational and experimental methods [11]. The influence of tube diameter and chirality on the stability of CNTs under electron and ion irradiation has been investigated using a combination of simulations and experiments. Various irradiations produce local damage to nanotubes, including crosslinking, which depends on the chirality of the nanotubes [12,13]. Ion irradiation can be employed for welding CNTs together, which is important for making a mechanically stable network of joined CNTs. A comparative study on various zero and one-dimensional nanoscale systems such as nanoclusters, nanowires, nanotubes and fullerenes was also carried out to understand the irradiation-induced effects [14,15]. The irradiation damage induces local structural changes and, in extreme circumstances, transformations that are dependent on the energy and the dose of the irradiating ion. The structural changes in the diameter and transformation of MWCNTs to amorphous carbon rods and wires under the higher fluence of ion irradiation are re-ported by various researchers [16–22]. Changes in the electronic structure, e.g., in the band gap of CNTs, have been studied due to the impact of energetic ions. The irradiation –induced defects can give rise to single and multiple peaks in the band gap of the semi-conducting nanotubes, and similar effects can be observed when several defects are close to each other [9,23]. Argon-ion irradiation has been used to form a tunnel barrier in MWCNTs and single quantum dots were fabricated [24]. The nature of electron and pro-ton irradiation damages and the morphological changes in single and multi-walled carbon nanotubes have also been extensively studied [25–29]. In MWCNTs, the multi-shell structure is much stiffer than the single-wall one, especially in compression [30]. MWCNTs are relatively easier to manufacture in large scale than SWCNTs [30,31]. Both types of carbon nanotubes are unique nano-structures for studying the mechanisms by which the energetic heavy ions interact with their constituents, including dissipate energy that breaks the individual and collective bonds. By choosing the appropriate structural dimensions for the single-and multi-walled carbon nanotubes, a comparative investigation of the irradiation induced damage to carbon nanotubes has been conducted to understand the dynamics of cluster sputtering and the associated multiple vacancy creation.

We conducted a series of experiments to investigate the pattern of heavy ion–induced fragmentation in carbon's nanostructures. These nanostructures include $C_{60}$ and a comparative study of MWCNT with graphite [32,33]. We are reporting the extension of the earlier study on SWCNTs in this paper. Cesium has been chosen, because its energy, intensity and dose can be accurately controlled and monitored in the source of negative ions with cesium sputtering (SNICS). The fragments are sputtered as neutrals and delivered as anions by the source. We can thus avoid the high temperature associated with plasma sources in such studies. Sputtered large car-bon clusters have higher probability of survival in the relative low temperatures associated with SNICS than those in hot plasmas. Mass spectrometry of the sputtered fragments as a function of $Cs^+$ energy and its dose identify the fragmenting species that can, in turn, provide information about the pattern of material loss and the stability of the nano-structures. The study aims to compare and contrast the fragmentation patterns and sputtering profiles of the irradiated nano-structures with those from the macro-structures of carbon. The rolled, single-shelled nanotubes used in the experiments reported here can also simulate the response of graphene sheets to energetic heavy ions. The design and construction of mono-layered graphene –for such experiments is much more difficult than that for SWCNTs. Irradiated graphite, i.e., the multi-layered graphene stack with macro dimensions, responds by ejecting $C_1$s and $C_2$s as the main sputtered species with much less intense $C_3$, $C_4$, $C_5$ and $C_6$ as well as larger clusters. These species indicate the creation of vacancies in the flat graphene sheets. The cross sections r for the emission of $C_1$, $C_2$ and higher clusters from graphite and multi-walled carbon nanotubes are a function of the energy and projectile type. There is also dose dependence as well [32].

Our present study shows that the cumulative damage in the irradiated single- and multi-walled carbon nanotubes shows similarities and subtle differences with each other and with the dam-age induced in graphite. This study focuses on the nature and extent of the $Cs^+$–induced structural damage in SWCNTs compared with that in MWCNTs. The identification tools are the sputtered carbon species $C_x$, x P 1 as a function of $Cs^+$ energy and the cumulative dose. Mass spectrometry of the sputtered anions for each energy step provides the existing and changing landscapes of the nanotubes that are being irradiated. From the mass spectra, the number densities of $C_x$ are plotted for consecutive Cs energies. The SEM micrographs and XRD data taken after the irradiation experiments are compared with those taken before, thus providing supporting evidence to recognize irradiation induced damage to single- and multi-walled nanotubes.

## 2. Experimental

We have chosen the source of negative ions with cesium sputtering (SNICS) that is designed to use $Cs^+$ irradiation of the chosen targets. Sputtering of the target species as monatomic, diatomic or higher clusters allows the experimenter to choose the desired species after mass analysis for acceleration in the 2 MV Pelletron as anions. The charge transfer in the stripping canal of high voltage terminal delivers positively charged ions from the accelerator. We utilized the source for studying Cs-induced fragmentation of single- and multi-walled carbon nanotubes. SWCNTs of 2 nm diameter and 3–13 lm length and MWCNTs of 10–20 nm diameter and 5–15 lm length were compressed in Cu bullets and used as targets for the source. The focus of the experiments included SWCNTs where the following two energy regimes were selected: one from 0.2 to 2.0 keV, with 0.2 keV steps, and the other, from 0.5 to 3.5 keV, with successively increasing energies in 0.5 keV steps. For MWCNTs, only the 0.5 to 3.5 keV range was selected with 0.5 keV steps. The source is unique for our experiments, as it is able to produce a stable $Cs^+$ beam with a wide range of current densities (from 100 lA to a few mA) for extended periods of time at a given energy in the range from 0.2–5.0 keV. The $Cs^+$ beam was focused on a 2–mm diameter area. The sputtered and recoiling atoms and clusters become negatively charged when leaving the target surface and were extracted from the source. The total energy of the extracted species was determined by the sum of the target and extraction voltages. The sputtered species were extracted at the constant energy of 30 keV. A momentum analyzer was used for the mass spectrometry of the sputtered species as a function of the cesium energy $E(Cs^+)$. Areas under the emitted cluster peaks were determined using the mass/charge (m/z) spectra as a function of $E(Cs^+)$. The data were calibrated against the $Cs^+$ number densities to evaluate the respective species' sputtering yields.

The experiments were conducted in two sequences. During the first sequence we irradiated a sample containing SWCNTS in the $Cs^+$ energy range of 0.2–2.0 keV in 19 energy steps. A second round of irradiations followed the first one in the same energy range. That implied a total of thirty-eight (38) sets of mass spectra for the two rounds of irradiations. The rationale for the experiments was to irradiate the pristine SWCNTs, starting with the lowest $Cs^+$ energy, so we could identify the nature of the sputtered species with minimum damage to the nanotubes. Even during this first round in which each spectrum was obtained in 300 s, the SWCNTs were ex-posed to an increasing energy and intensity $Cs^+$ beam for a total irradiation time 6,000 s. The second round of irradiations at the same energies revealed the nature and extent of the damage, mostly as a function of $E(Cs^+)$, as discussed in the next section.

To study and monitor the structural changes associated with the energetic heavy ion irradiation dose another series of experiments was performed with fresh samples. The first sequence of irradiations was performed with $E(Cs^+) = 0.5–3.5$ keV, i.e., the respective mass spectra obtained using the 0.5 keV step. Next the SWCNTs were irradiated continuously at 5.0 keV for 1,800 s. Total doses of 3 $10^{16}$ $Cs^+$ cm 2 were implanted at 5.0 keV. After this round of irradiations, the sequence of mass spectra as a function of $Cs^+$ energy was repeated in the range $E(Cs^+) = 0.5–3.5$ keV. For this sample of SWCNTs, SEM micrographs were collected before and after the irradiation sequences to compare and contrast the radiation–induced effects on the structure of single-walled carbon nanotubes. The results are presented and discussed in later sections. MWCNTs were subjected to similar irradiation sequences under similar conditions to study any contrasts. To compare the irradiation effects in the two structures, we compressed MWCNTs of 10–20 nm diameter and 5–15 lm length in the Cu targets of SNICS, and mass spectra with $E(Cs^+)$ = 0.5–3.5 keV were obtained. A round of continuous irradiations at 5.0 keV for 5,000 s was followed by a repeat of the earlier measurements. The comparison of the cluster yields from SWCNTs and MWCNTs under similar $Cs^+$ irradiation conditions was helpful to us in understanding the nanotube fragmentation mechanisms that leave behind multiple vacancies in the damaged nanotube structures.

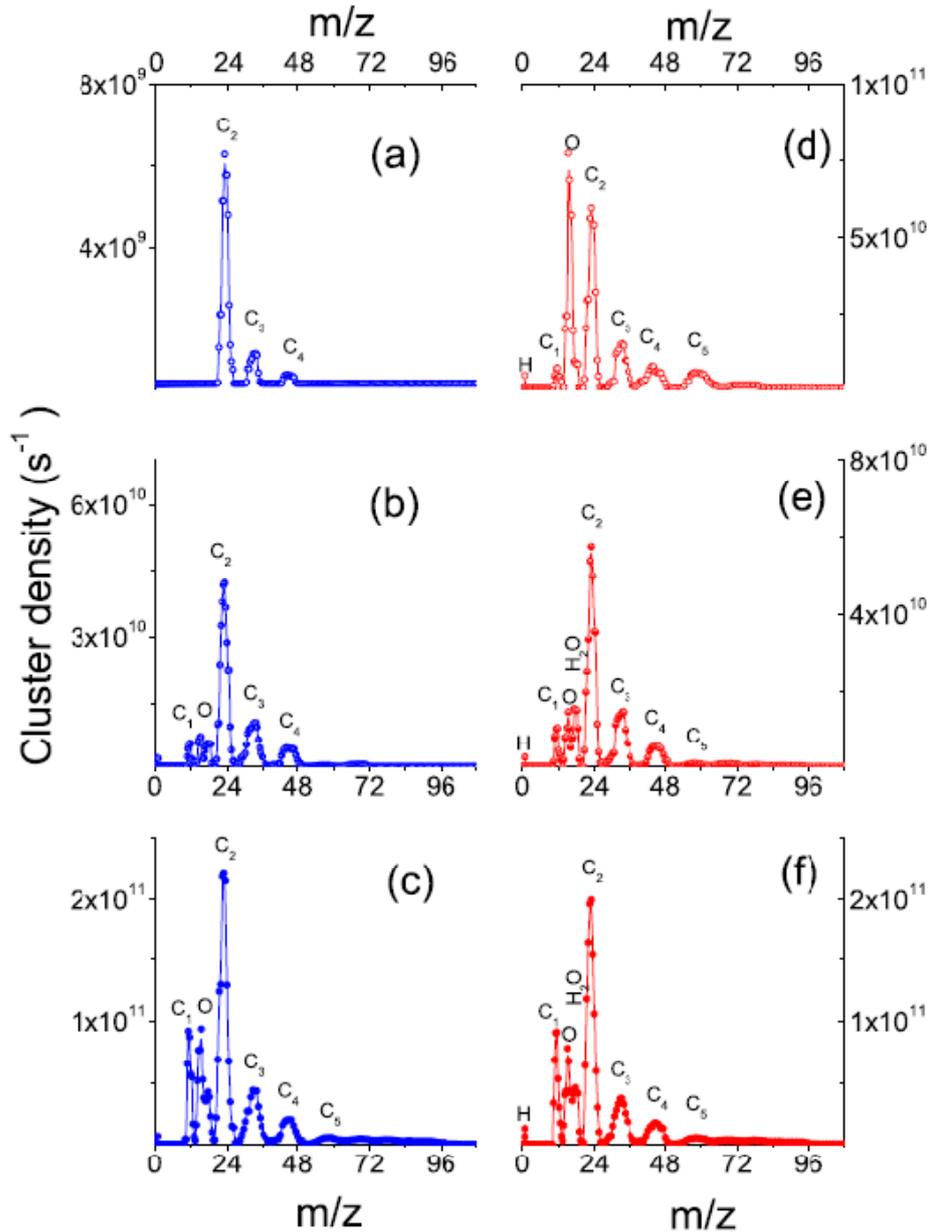

Fig. 1. Six mass spectra are plotted for three $Cs^+$ energies: 0.2, 0.8 and 2.0 keV (a–c) show the pristine SWCNTs, whereas (d–f) show the Cs-rich, irradiated ones.

In SNICS, the target surfaces are covered with one or more layers of Cs atoms. The sputtered species penetrate this Cs layer and acquire electrons from Cs atoms. Higher sputtering yields for most solids coupled with high electron transfer efficiency are the hall-mark of this source. The sputtering output from SNICS takes the form of anions. Di-anions or higher negatively charged species are possible but occur with very low probability [35]. $C_2^{2-}$, for example, should appear in the m/z spectra overlapping with $C_1$ ; however, we do not observe even $C_1$ at low $Cs^+$ energies. Multiply charged anions have been investigated with special designs [36].

## 3. SWCNTs irradiated with low E(Cs$^+$) and dose

Fig. 1 shows two sets of three mass spectra each. Figs. 1(a–c) are the m/z spectra from the pristine SWCNTs at $E(Cs^+)$ = 0.2, 0.8 and 2.0 keV, respectively. These 3 spectra are taken out of the consecutive series of 19 spectra. The second set is shown in Fig. 1(d–f) at the same $Cs^+$ energies but from the repeat sequence of the irradiations after the first round.

The second set can be considered to originate from a Cs-rich, irradiated set of SWCNTs. The first spectrum (Fig. 1(a)) shows only three species, i.e., $C_2$, $C_3$ and $C_4$, with successively reducing number densities. The most conspicuous species by its absence is $C_1$. At 0.8 keV, $C_1$ appears with its more intense, higher C clusters. The minimum threshold for the $Cs^+$ energy $E_T(Cs^+)$ is 0.4 keV, the energy at which $C_1$ is first ob-served in the sputtered species' mass spectrum. In Fig. 1(b), in addition to $C_1$, $C_2$, $C_3$, $C_4$, $C_5$ and $C_6$ are also present at 0.8 keV but with an order of magnitude smaller densities. At $E(Cs^+)$ = 2.0 - keV, heavier carbon clusters $C_x$ with x up to 9 can be identified and, merged above a broad spectrum. At higher energies, all clusters are emitted, including even the Cs-substituted carbon clusters, also re-ported from MWCNTs [32]. The sputtering of the Cs-substituted carbon clusters, e g., Cs–$C_2$ is an indication of the deposition of large number densities of Cs a few nm below the surface. The mass spectra from the heavily irradiated and Cs-rich SWNTS show characteristic differences to those obtained from the pristine set. Fig. 1(d), as opposed to 1(a), has the entire range from $C_1$ to $C_7$, showing the effect of earlier irradiations. Similar differences can be identified in Figs. 1(e and f). The conspicuous presence of the $H_2O$, O and H peaks are a reminder of the water impurity, which has proved to be difficult to remove. We annealed our samples but were unable to remove the adsorbed water completely. A separate study on water as an impurity in nanotubes will be reported elsewhere [34]. It can be considered an indicator of the state of nanotube damage. For example, the differences in Figs. 1(a and d) show not only the carbon cluster type and their relative densities, but they also represent differences of the water-related peaks.

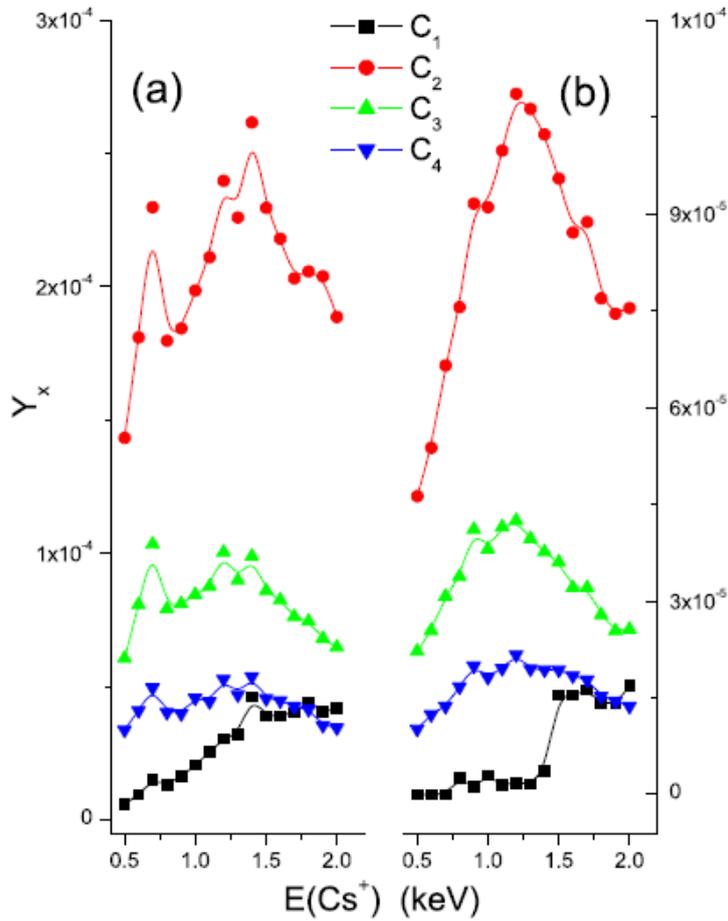

Fig. 2. Sputtering yields of $C_1$, $C_2$, $C_3$ and $C_4$ are plotted from the pristine and Cs– irradiated sample of SWCNTs in (a and b), respectively

Fig. 2 presents the sputtering yield for each of the sputtered species $Y_x = I_x/I(Cs^+)$ obtained from the cumulative data of 38 mass spectra. Here the current densities $I_x$ of each cluster $C_x$ at successive energy steps are normalized to $Cs^+$ current densities $I(Cs^+)$. We have plotted the sputtering yield $Y_x$; x = 1–4, as a function of $E(Cs^+)$. Fig. 2(a) is for the pristine SWCNTs, and Fig. 2(b) shows the heavily irradiated, Cs-rich SWCNTs. Fig. 2(a) shows the four selected species appearing in the sputtering yield versus $Cs^+$ energy. These appear in the decreasing order, $C_2$, $C_3$, $C_4$ and $C_1$. $C_2$ is the most abundantly sputtered species throughout the $Cs^+$ energy variation. There is a broad peak for $C_2$, $C_3$ and $C_4$ between $E(Cs^+)$ = 1.0– 1.5 keV. The yield of $C_1$, on the other hand, rises almost linearly with the sputtering ion's energy. Fig. 2(b) has the repeat mass spectra converted to the sputtering yields of the respective species that show a sharp decrease between 0.2 and 0.5 keV. The decrease has a large slope for $C_2$, $C_3$ and $C_4$. Between 0.5 and 3.5 keV, the three species show broad spectral peaks, similar to those in Fig. 2(a). $C_1$ is not sputtered from the already damaged nanotubes until the energy of the irradiating $Cs^+$ reaches 1.5 keV, after which it makes a steady contribution.

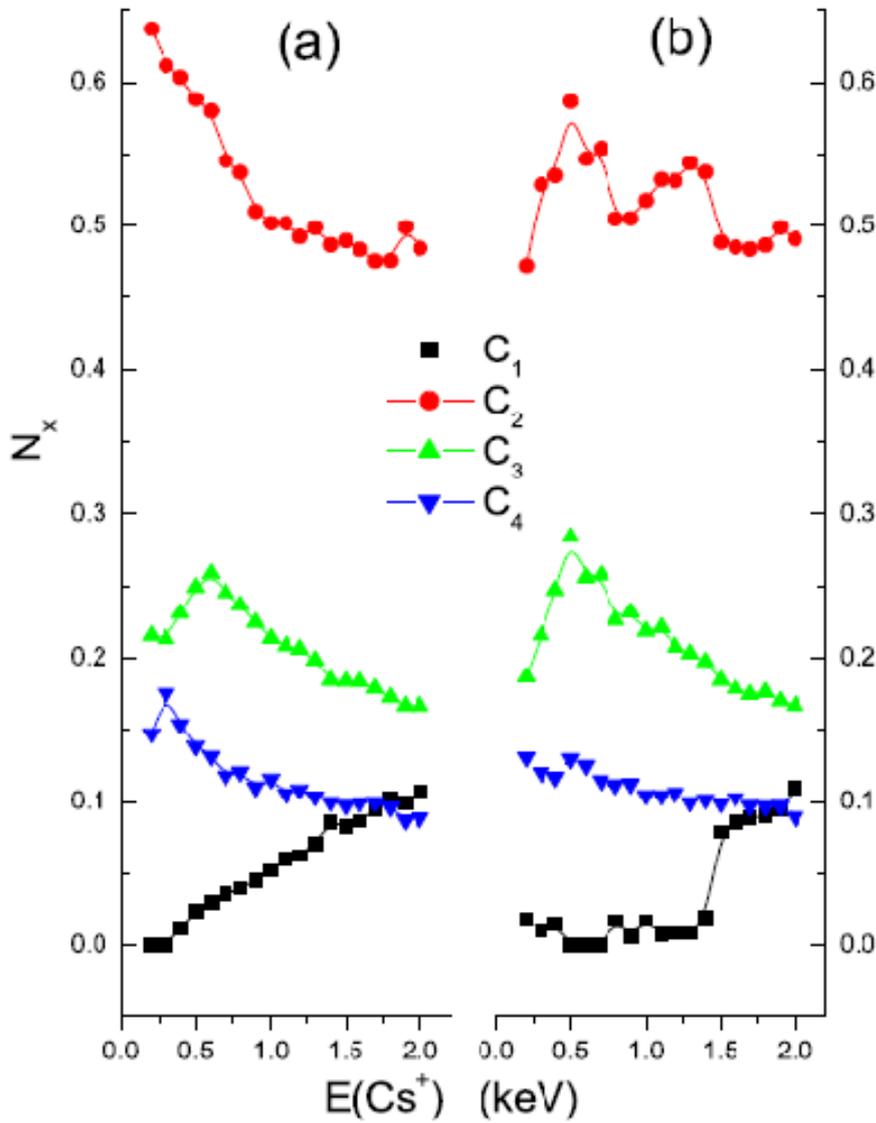

Fig. 3. Normalized yields of the four sputtered species is calculated from the respective sputtering yields and plotted as a function of E(Cs$^+$) for the pristine (a) and the irradiated SWCNTs (b).

The sputtering yield data for $C_1$, $C_2$, $C_3$ and $C_4$ presented in Figure 2 may provide clues to the mechanisms of fragmentation and the consequent multiple vacancy generation in the SWCNTs. Yet another way of understanding the dynamics of defect generation is shown in Fig. 3. The normalized relative yield of the clusters $C_x$ emitted from the irradiated nanotubes can be shown as $N_x = Y_x/ RY_x$, a function of Cs$^+$ energy. Fig. 3(a) is from the pristine SWCNTs, and 3(b) shows the heavily irradiated SWCNTs. Both figures show that more than half of all emissions from the fragmenting nano-tubes are $C_2$s. That implies the generation of double vacancies as the dominant defect resulting from energetic irradiations at all energies. The next most probable fragmenting route is via the emission of a $C_3$, and thus by the creation of triple vacancies, again at all energies and from the pristine and heavily irradiated nanotubes.

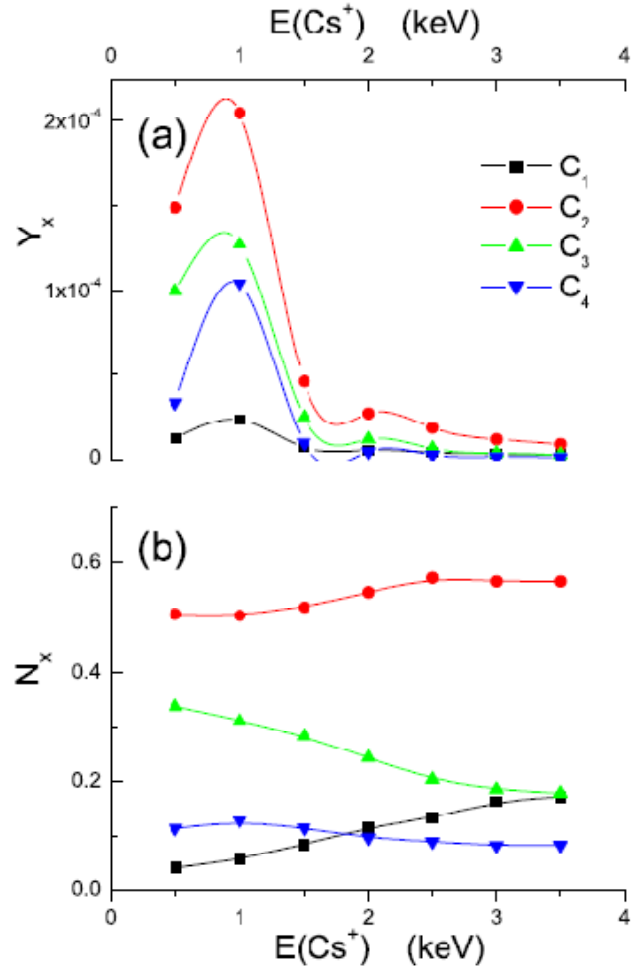

Fig. 4. Sputtering yield $Y_x$ in (a) and the normalized yield $N_x$ in (b) are plotted for $C_1$, $C_2$, $C_3$ and $C_4$ against $E(Cs^+)$ for the pristine SWCNTs

The normalized yield $N_x(=Y_x/RY_x)$ plots provide us the landscape of the comparative densities. $C_2$, $C_3$ and $C_4$ are the main species emitted from the irradiated SWCNTs, with subtle differences between their number densities in the pristine set of nanotubes and those in the heavily irradiated set. These differences illustrate the nature of the damage to SWCNTs and the consequent emergence of the sputtered $C_x$-initiated structures. Fig. 3(a) shows $C_2$'s relative yield decreasing from 64% at $E(Cs^+)$ = 0.2 keV to 50% between 1 and 2 keV. $C_3$'s yield increases from 20% to a broad peak of 25% at 0.6 keV and stabilizes around its initial yield. $C_4$ is a stable species with 10% relative yield for the entire $E(Cs^+)$ range. The relative population of $C_1$ increases from 0 to 10%, indicating a uniform increase in the production of single vacancies with $E(Cs^+)$. Figure 3(b) shows a similar broad feature of the cluster emission profile that is obtained after the SWCNTs are irradiated for 1,800 s. $C_2$, $C_3$ and $C_4$ are the main constituents emitted from the $Cs^+$ sputtered nanotubes. The $C_2$ contribution to the total normalized yield shows fluctuations of approximately 55%. $C_3$'s yield is similar to that shown in Fig. 3(a). $C_4$ also has a steady 10%. $C_1$ is not sputtered from the heavily irradiated ensemble of SWCNTs until $E(Cs^+)$ 1.5 keV, after which its share is steady at approximately 10%, indicating that $C_1$ may be a by-product of the larger $C_x$ fragmentation.

## 4. Higher E(Cs⁺) and dose irradiations of SWCNTs

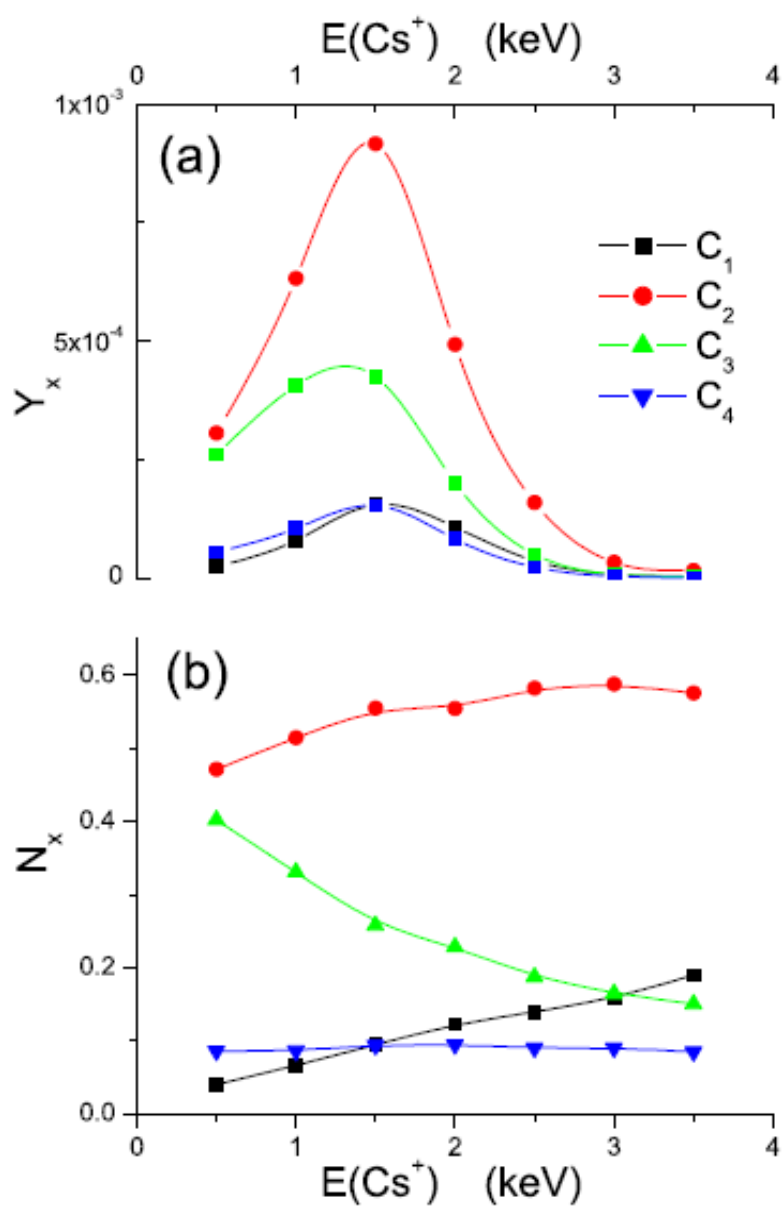

Fig. 5. Data similar to those in figure 4 for $C_1$–$C_4$ are shown here from the heavily irradiated SWCNTs.

Another fresh sample of the compressed SWCNTs was subjected to 0.5–3.5 keV $Cs^+$ beam irradiations. The energy step was 0.5 keV. This larger energy step was chosen to span a larger energy range to allow monitoring the variations of the sputtered species' sputtering yields as a function of energy without subjecting SWCNTs to higher $Cs^+$ yields as performed in the first sequence of experiments. Fig. 4(a) has the sputtering yield $Y_x$ as a function of irradiating $Cs^+$ energy $E(Cs^+)$. The peaks in the respective $C_x$ yield are approximately 1 keV for all species, including $C_1$. These data are important because we may be able to separate the effects of the irradiating ion energy from effects occurring as a result of the cumulative dose-dependent damage. The yields have a sharp decrease between 1 and 2 keV. Fig. 4(b) shows the normalized yield $N_x(=Y_x/RY_x)$ as a function of $E(Cs^+)$. The $C_2$ relative yield is almost constant, with a small increase at higher $Cs^+$ energies. This confirms the observation from the low–energy experiments reported earlier, stating that $C_2$ is the main fragmenting species with the subsequent double vacancy creation. The normalized yield shows a slight increase as opposed to the decrease after $E(Cs^+)$ P 1 keV, ob-served in Fig 2. $C_3$ shows the peak at approximately 1 keV, with a gradual decrease but without the sharp features observed in Fig. 2(a) and 3(a). $C_4$ has a steady share of the normalized yields at approximately 10%, whereas $C_1$ increases linearly from 2 to 20%.

Next 1,800 s continuous irradiation occurs with 5 keV $Cs^+$ beam implants 3 $10^{18}$ $Cs^+$ cm $^2$, with considerable damage to the SWCNTs. Fig. 5 (a) has the sputtering yields for the four species. It shows a clear broad peak 1.5 keV in the spectra of all species. The normalized yields of all species show approximately the same pattern as shown in Fig. 4(b). Here, $C_2$ shows a slight increase; $C_3$ decreases; $C_4$ remains steady, and $C_1$ shows an increasing trend that is almost similar to the one in the sputtering yields from the un-irradiated pristine samples. The shift of the peaks in the sputtering yields of all species from $E(Cs^+)$ 1.0 keV from the un-irradiated to $E(Cs^+)$ 1.5 keV for the heavily irradiated, damaged, and $Cs^+$-rich SWCNTs is the most significant indicator of the irradiation damage. All data points have been connected using by spline or B-spline interpolation directly from the Origin software. Spline interpolation uses low-degree polynomials in each of the intervals such that the connecting curve is smooth. These interpolations serve simply as guides to the eye and do not represent the theoretical results.

## 5. Comparison with the irradiated MWCNTs

MWCNTs were subjected to a similar round of irradiations as was performed for SWCNTs. The results allow us to compare the $Cs^+$ induced fragmentations in the two types of nanotubes as a function of the same range of $E(Cs^+)$. Comparing the sputtering yields $Y_x$ and the normalized yields $N_x$ from Figs. 4 and 6 for the two types of nanotubes shows the following:

1. Both types of carbon nanotubes seem to fragment and have $C_2$ as the dominant sputtered cluster with the consequent creation of di-vacancies for the entire range of $Cs^+$ energies.

2. $C_3$ is the next highest density cluster to be sputtered at almost all irradiation energies.

3. The sputtering yields are about two orders of magnitude higher for SWCNTs.

4. For SWCNTs, there are well –defined peaks 1 keV for the number densities of the three major sputtered clusters, i.e., $C_2$, $C_3$ and $C_4$.

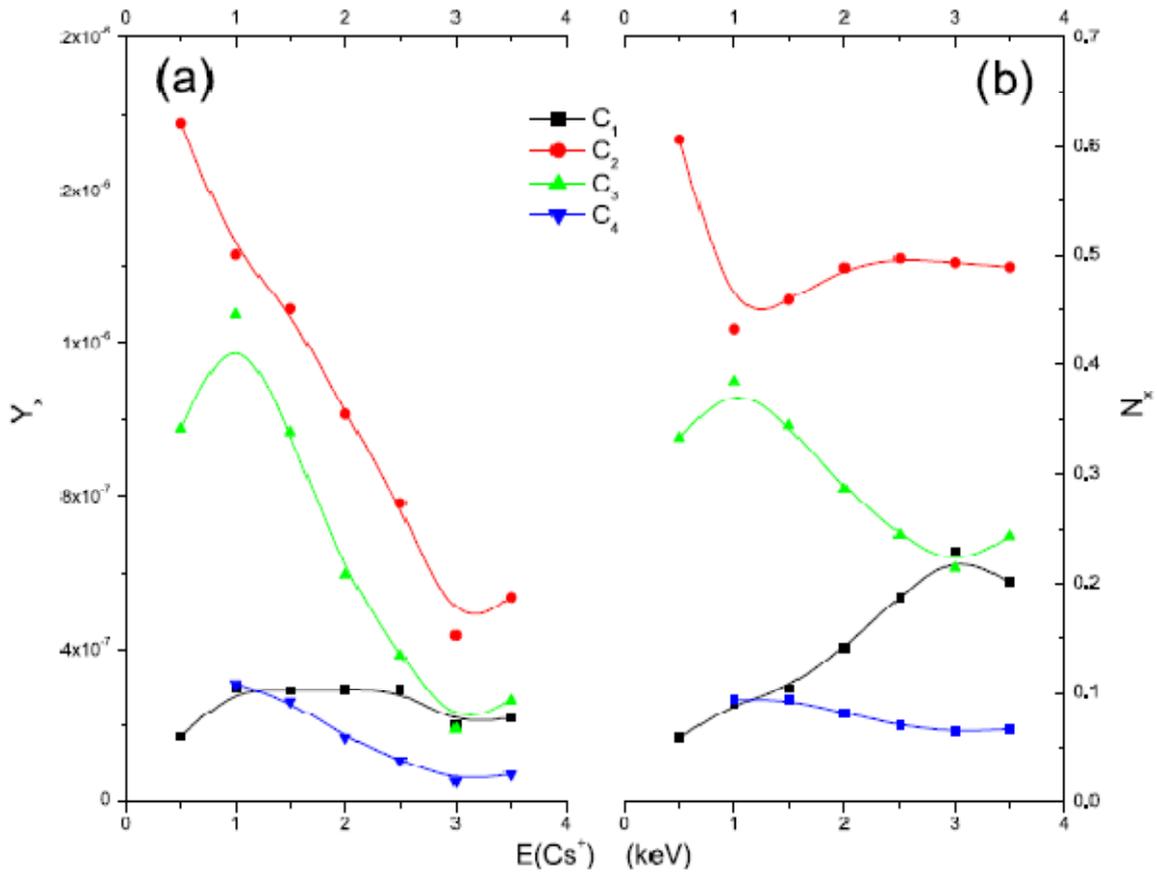

Fig. 6. Sputtering $Y_x$ and normalized $N_x$ yields for $C_1$–$C_4$ are plotted for the MWCNTs in the $Cs^+$ energy range $E(Cs^+)$ = 0.5–3.5 keV.

5. For MWCNTs there is a general linearly decreasing trend for $C_2$ and $C_3$. $C_4$ and $C_1$ have low puttering yields and do not show variations

6. The normalized yields $N_x$ versus $E(Cs^+)$ show the relative cluster densities. Here, $C_2$'s relative yield $50 \pm 4\%$ for all emissions throughout the range of the irradiating ion.

7. $C_3$'s relative share of the sputtered species is between 35% at 0.5 keV and reduces to 20% at 3.5 keV.

8. $C_4$ appears as a stable species, with a 10% share at low $E(Cs^+)$=0.5 keV that reduces gradually to 7–8% at higher $Cs^+$ energies.

9. $C_1$ shows a linear increase in both cases, going from 5% to 20% between $E(Cs^+)$=0.5 and 3.5 keV.

# 5. Secondary electron microscopy and XRD of the pristine and heavily irradiated SWCNTs and MWCNTs

Fig. 7 shows the SEM micrograph of the pristine and heavily irradiated samples of SWCNTs; (a) shows pristine sample, whereas (b) and (c) are the further magnified images of the selected portion.

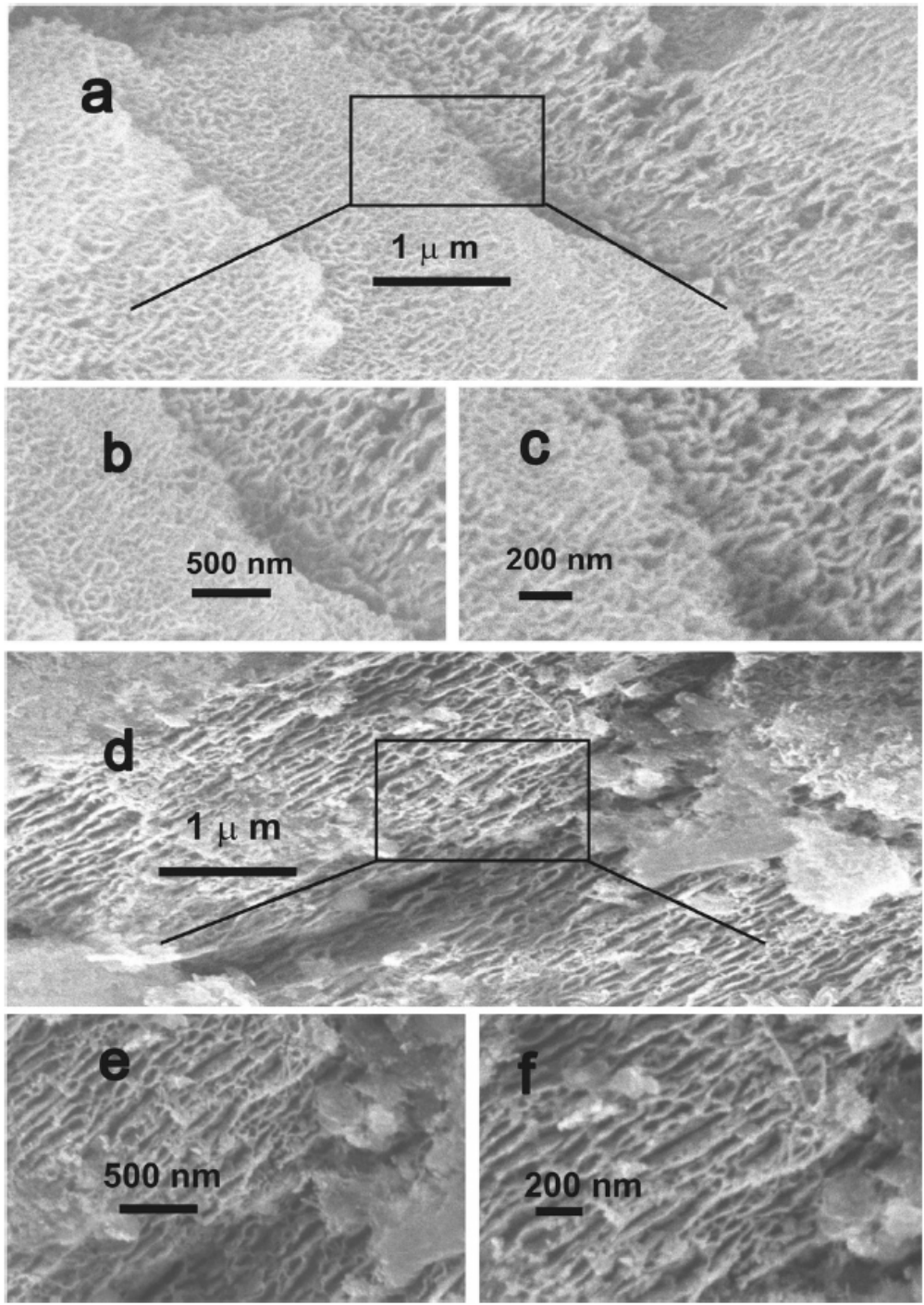

Fig. 7. SEM micrographs of SWCNTs (a) are for the pristine sample; (b) and (c) show the same un-irradiated portion with larger magnification. Part (d) shows the heavily irradiated, $Cs^+$ rich sample; magnified images are shown in (e) and (f).

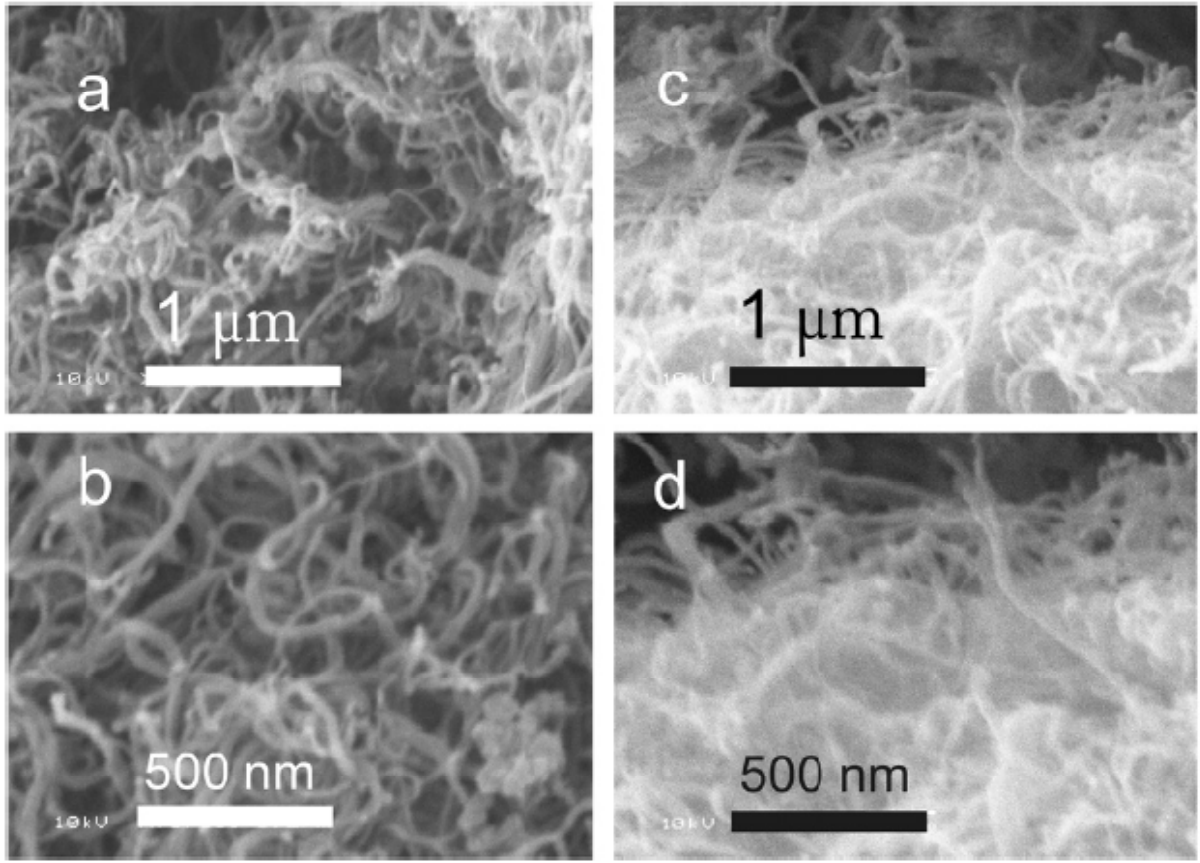

Fig. 8. SEM micrographs of MWCNTs (a) and (b) are for the pristine sample at different magnifications; (c) and (d) are for the heavily irradiated samples.

Fig. 8 shows the SEM micrograph of the pristine and the heavily irradiated, $Cs^+$ rich samples of MWCNTs. Fig. 8(a) shows the pristine sample at lower magnification, and 8(b) shows the sample at higher magnification. From both figures one can see that there are empty spaces between the pristine MWCNT bundles. Parts (c) and (d) show the micrographs of MWCNTs that were heavily bombarded by energetic $Cs^+$ ions, and the maximum damage is ob-served in this region. One can identify the filling of the open spaces. Cylindrical tubes have been heavily damaged with the sub-sequent sputtering of the radical carbon atoms and clusters may be acting as fillers and bonding agents.

Fig. 9 shows the XRD spectra from the pristine and heavily irradiated samples of SWCNTs and MWCNTs. In both cases, the intensity of the peak corresponding to the main graphitic planes (002) at 26.2L –decreased after irradiation and became broader, signifying the partial loss of the crystal structure. The substrate peaks are not shown in the irradiated samples because several layers of irradiated carbon nanotubes were taken out from the holder and tapped on glass slide, XRD was then performed to avoid substrate peaks. In MWCNT spectra, the peak at 22.5º is from tape on the glass slide.

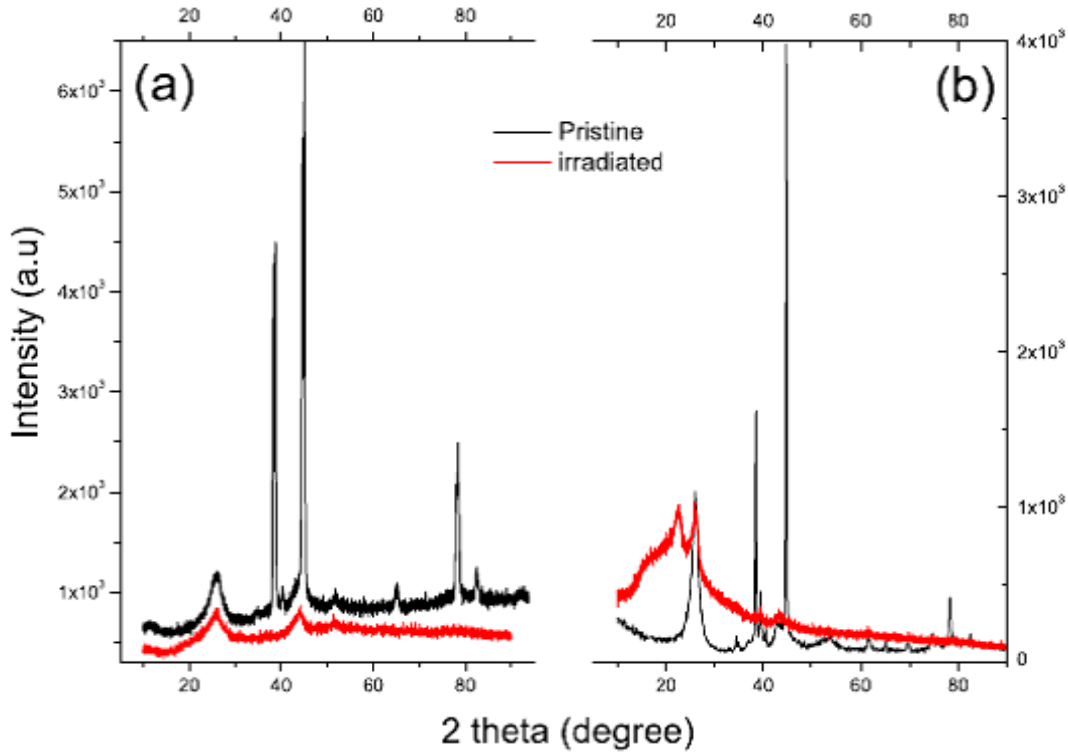

Fig. 9. XRD spectra for (a) SWNT and (b) MWNT before and after irradiation with $Cs^+$.

# 7. Conclusions

An investigation into the $Cs^+$-induced fragmentation resulting in the generation of multiple vacancies in the pristine and heavily irradiated SWCNTs as a function of energy and $Cs^+$ dose has been performed, and the results were compared with those obtained under similar conditions from MWCNTs. $C_2$, $C_3$ and $C_4$ are the most persistent sputtered species from the SWCNTs, and $C_1$ is the least intense. We interpret the dominance of the cluster emission over that of the monomers as the ease with which multiple vacancies can be created as opposed to single vacancies. The mass spectra from the heavily irradiated SWCNTs show characteristic differences with those obtained from pristine samples. From heavily irradiated samples, an entire range of sputtered clusters from $C_1$ to $C_7$ is observed. $C_2$ proved to be the most dominant species, followed by $C_3$, $C_4$, $C_1$ and the higher clusters. For the lower energy range of $Cs^+$, the maximum yield of $C_2$, $C_3$ and $C_4$ is obtained between $E(Cs^+)$ 1.0 and 1.5 keV for both the pristine and $Cs^+$ rich samples. $C_1$ is not sputtered from already damaged nanotubes until the energy of irradiating $Cs^+$ reaches 1.5 keV, indicating that $C_1$ is a by-product of the larger clusters' fragmentation. $C_1$'s yield in-creases linearly with $E(Cs^+)$ for the pristine sample. The normalized relative yields of clusters show that more than half of all emissions from the pristine and heavily irradiated nanotubes are $C_2$s. The generation of double vacancies, followed by triple vacancies, emerges as the dominant defects that occur in the heavy ion–irradiated carbon nanotubes.

All clusters, including $C_1$, have maximum yields around 1.0 keV for pristine SWCNTs samples and a broad peak around 1.5 keV for heavily irradiated samples. For MWCNTs, the yields of the sputtered clusters show that $C_2$ is followed by $C_3$ as the most intense sputtered species, similar to the results obtained from SWCNTs under same conditions. Highly irradiated carbon nanostructures are internally massively deformed, and the sputtered species may accumulate and form connecting bridges between the nanotubes. The sputtered species, after fragmenting from nanotubes and accumulating in the intra-single and multi-walled carbon nanotubes spaces might be responsible for the merged and perhaps welded-looking structures in the SEM micrographs.

# References

[1] S. Iijima, Nature 354 (1991) 56–58.
[2] T.W. Ebbesen, P.M. Ajayan, Nature 358 (1991) 220–222.
[3] M.S. Dresselhaus, G. Desselhaus, P.C. Eklund, Science of Fullerenes and Carbon Nanotubes, Academic Press, London, 1996.
[4] P.J.F. Harris, Carbon Nanotube Science. Synthesis, Cambridge University Press, Cambridge, Properties and Applications, 2011.
[5] J. Pomoell, A.V. Krasheninnikov, K. Nordlund, J. Keinonen, J. Appl. Phys. 96 (2004) 2864–2871.
[6] J. Pomoell, A.V. Krasheninnikov, K. Nordlund, J. Keinonen, Nucl. Instrum. Meth. Phys. Res B 206 (2004) 18–21.
[7] O. Lehtinen, T. Nikitin, A.V. Krasheninnikov, L. Sun, F. Banhart, L. Khriachtchev, J. Keinonen, New J. Phys. 13 (2011) 073004.
[8] A.V. Krasheninnikov, K. Nordlund, J. Keinonen, Phys. Rev. B 65 (2002) 165423– 165431.
[9] A. Tolvanen, J. Kotakoski, A.V. Krasheninnikov, K. Nordlund, Appl. Phys. Lett. 91 (2007) 173109–173111.
[10] A.V. Krasheninnikov, K. Nordlund, Keinonen, Phys. Rev. B 66 (2002) 245403– 245409.
[11] B. Ni, R. Andrews, D. Jacques, D. Qian, M.B.J. Wijesundara, Y. Choi, L. Hanley, S.B. Sinnott, J. Phys. Chem. B 105 (2001) 12719–12725.
[12] S.K. Pregler, S.B. Sinnott, Phys. Rev. B 73 (2006) 224106–224109.
[13] Z. Xu, W. Zhang, Z. Zhu, C. Ren, Y. Li, P. Huai, J. Appl. Phys. 106 (2009) 043501– 043514.
[14] A.V. Krasheninnikov, K. Nordlund, Nucl. Instrum. Meth. Phys. Res. B 216 (2004) 355–366.
[15] A.V. Krasheninnikov, K. Nordlund, Appl. Phys. 107 (2010) 071301–071371.
[16] K. Dharamvir, K. Jeet, C. Du, N. Pan, V.K. Jindal, J. Nano Res. 10 (2010) 1–9.
[17] M.S. Raghuveer, P.G. Ganesan, J.D. Gall, G. Ramanath, M. Marshall, I. Petrov, Appl. Phys. Lett. 84 (2004) 4484–4487.
[18] B.Q. Wei, J.D. Gall, P.M. Ajayan, G. Ramanath, Appl. Phys. Lett. 83 (2003) 3581– 3583.
[19] O. Lehtinen, L. Sun, T. Nikitin, A.V. Krasheninniko, L. Khriachtchev, J.A.R. Manzo, M. Terrones, F. Banhart, J. Keinonen, Physica E 40 (2008) 2618–2621.
[20] Z. Ni, Q. Li, J. Gong, D. Zhu, Z. Zhu, Nucl. Instrum. Meth. Phys. Res B 260 (2007) 542–546.
[21] H.M. Kim, H.S. Kim, S.K. Park, J. Joo, T.J. Lee, C.J. Lee, J. Appl. Phys. 97 (2005) 026103–026106.
[22] A.V. Krasheninnikov, K. Nordlund, M. Sirviö, E. Salonen, J. Keinonen, Phys. Rev. B 63 (2001) 245405–245408.
[23] A. Tolvanen, G. Buchs, P. Ruffieux, P. Gröning, O. Gröning, A.V. Krasheninnikov, Phys. Rev. B 79 (2009) 125430–125442.
[24] M. Suzuki, K. Ishibashi, K. Toratani, D. Tsuya, Y. Aoyagi, Appl. Phys. Lett. 81 (2002) 2273–2275.
[25] Y. Gan, J. Kotakoski, A.V. Krasheninnikov, K. Nordlund, F. Banhart, New J. Phys. 10 (2008) 023022–023030.
[26] K. Mølhave, S.B. Gudnason, A.T. Pedersen, C.H. Clausen, A. Horsewell, P. Bøggild, Ultramicroscopy 108 (2007) 52–57.

[27]   H.C. Kiang, W.A. Goddard, R. Beyers, D.S. Bethune, J. Phys. Chem. 100 (1996) 3749–3752.
[28]   B. Khare, M. Meyyappan, M.H. Moore, P. Wilhite, H. Imanaka, B. Chen, Nano Lett. 3 (2003) 643–646.
[29]   V.A. Basiuk, K. Kobayashi, T.K.Y. Negishi, E.V. Basiuk, J.M.S. Blesa, Nano Lett. 2 (2002) 789–791.
[30]   F. Béguin et al., in Carbon Nanomaterials, ed Y. Gogotsi, CRC Press, Boca Raton, 2006.
[31]   P. Barone et al., In Carbon Nanotubes Properties and Application, ed, CRC Press, Boca Raton, M.J. O'Connell, 2006.
[32]   S. Ahmad, Nucl. Instrum. Meth. Phys. Res B 271 (2012) 55–60.
[33]   S. Zeeshan, S. Javeed, S. Ahmad, Int. J. Mass Spectrom. 311 (2012) 1–6.
[34]   I. Aslam, S. Ahmad to be published.
[35]   R. Middleton, Nucl. Instrum. Meth. 144 (1977) 373–399.
[36]   S.N. Schauer, P. Williams, Phys. Rev. Lett. 65 (1990) 625–628.